\documentclass{aa}
\usepackage{graphics}
\input psfig.sty
\def\hi{\hbox{\ion{H}{i}}~}
\def\halpha{\hbox{H{$\alpha$}}~}
\def \kmsws {$\rm km~s^{-1}$}
\def \psqcm {$\rm cm^{-2}$~}

\begin{document}

   \thesaurus{04     
              (09.03.1;  
               09.19.1;  
               10.08.1;  
	       10.19.3)  
               }
   \title{The head-tail structure of high-velocity clouds}

   \subtitle{A survey of the northern sky}

   \author{C. Br\"uns\inst{1,2}
         \and J. Kerp\inst{1}
         \and P.M.W. Kalberla\inst{1}
         \and U. Mebold\inst{1}
          }

   \offprints{C. Br\"uns, Bonn address}

   \institute{Radioastronomisches Institut der Universit\"at Bonn, 
              Auf dem H\"ugel 71, D-53121 Bonn, Germany\\
              email: cbruens@astro.uni-bonn.de
         \and
   The Australia Telescope National Facility, CSIRO, 
       PO Box 76, Epping NSW 1710, Australia\\
             }

   \date{Received  25th October, 1999; accepted 8th February 2000 }

   \maketitle

   \begin{abstract}

   We present new observational results on high-velocity clouds (HVCs)
   based on an analysis of the Leiden/Dwingeloo \hi survey. We cataloged all
   HVCs with  N$_{\rm HI}\geq 1\cdot$10$^{19}\,{\rm cm^{-2}}$ and
   found 252 clouds that form a representative flux limited sample. The detailed
   analysis of each individual HVC in this sample revealed a significant
   number of HVCs (nearly 20\%) having simultaneously a velocity and a column 
   density gradient. These HVCs have a cometary appearance in the 
   position-velocity representation and are called henceforward 
   head-tail HVCs (HT HVCs).
   The head is the region with the highest column density of the HVC, while the
   column density of the tail is in general much lower (by a factor of 2 - 4). 
   The absolute majority of the cataloged HVCs belongs to the well known HVC
   complexes. With exception of the very faint HVC complex L, all HVC complexes
   contain HT HVCs. The HT HVCs were analyzed
   statistically with respect to their physical parameters like position, 
   velocity ($v_{\rm LSR},~v_{\rm GSR}$), and
   column density. We found a linear correlation between the fraction of 
   HVCs having a head-tail structure and the peak column density of the HVCs. 
   While there is no correlation between the fraction of HT HVCs and v$_{\rm
   LSR}$, we found a dependence of the fraction of HT HVCs and v$_{\rm GSR}$.
   There is no significant correlation between the
   fraction of HT HVCs and the parameters galactic longitude and latitude.
   The HT HVCs may be interpreted as HVCs that are
   currently interacting with their ambient medium. In the context of 
   this model the tails represent material that is stripped off from
   the HVC core. We discuss the implications of this model for galactic and 
   extragalactic HVCs.

   \end{abstract}

%

\section{Introduction}
   
HVCs -- first discovered by Muller et al. (\cite{muller}) -- are defined as
neutral atomic hydrogen clouds with unusual high radial velocities (relative 
to the local--standard--of--rest frame, LSR) which deviate significantly from a simple
galactic rotation model.

Despite 36 years of eager investigations there is no general consensus
on the origin and the basic physical parameters of HVCs.
This is mainly because HVCs are difficult to detect in emission other than 
\hi 21-cm line radiation. HVCs mostly appear as ``pure'' neutral atomic hydrogen 
clouds. Absorption line studies provide information on the ionization state 
and the metalicity of HVCs. Their results indicate that the bright and very 
extended HVC complexes consist (at least partly) of processed material, having 
$\le$ 1/3 of the solar abundances (see Wakker \& van Woerden \cite{wakker-rev} 
for a recent review). Recent observational results present evidence for 
emission of ionized atoms associated with HVCs.
Tufte et al. (\cite{tufte}) for instance presented \halpha
emission associated with the HVC complexes M, A and C.
Also the Magellanic Stream was found to be bright in the \halpha line emission
(Weiner \& Williams \cite{weiner}).
In the soft X-ray regime evidence is presented that the 
HVC complexes M, C, D and GCN are associated with excess soft X-ray radiation, 
produced by a plasma of a temperature $T_{\rm plasma} = 10^{6.2}$ K 
(Herbstmeier et al. \cite{herbstmeier}, Kerp et al. \cite{kerp}).
Also in the $\gamma$-ray regime the detection of excess 
$\gamma$-ray emission is claimed towards the HVC complex M (Blom et al. 
\cite{blom}).

However, the most critical issue of HVC research is the distance uncertainty
to the HVCs. 
Danly et al. (\cite{danly}), Keenan et al. (\cite{keenan}) and Ryans
et al. (\cite{ryans}) consistently determined an upper distance limit of 
z $\leq$  3.5 kpc to HVC complex M.
The most important step forward is the very recently determined distance bracket
of 2.5 $\leq$ z $\leq$ 7 kpc
towards HVC complex A by van Woerden et al. (\cite{van Woerden}).
These results clearly indicate that the HVC complexes M and A belong to the 
Milky Way and its gaseous halo.

Parallel to the growing evidence that a significant fraction of the HVC 
complexes belong to the Milky Way and its halo, Blitz et al. (\cite{blitz}) 
supported the hypothesis that some HVCs are of extragalactic origin.
They argued, that it is reasonable to assume that primordial gas -- left
over from the formation of the Local Group galaxies -- may appear as HVCs.
Observational evidence for such a kind of HVC may be found by the detection of
the highly ionized high-velocity gas clouds by Sembach et al. (\cite{sembach}),
 because of its very low pressure of about 5 K ${\rm km~s^{-1}}$. 

The Magellanic System is a special case outside both major approaches. 
The Magellanic Stream (MS) and the Leading Arm (LA) (Putman et al.
\cite{putman}) both form coherent structures over several tens of degrees 
having radial velocities in the HVC regime. Their gas represents most likely 
the debris caused by the tidal interaction of the Magellanic Clouds with the 
Galaxy at distances of tens of kpc. 

The physical conditions of the HVCs located in the gaseous Galactic halo or in
the intergalactic space as well as their chemical compositions should be 
significantly different.
HVCs located in the intergalactic space are only exposed to the extragalactic 
radiation field.
It is reasonable to assume that their column density distribution is dominantly
modified by gravitational forces of the Local Group galaxies.
In contrast, the gaseous distribution of the HVCs located within the
neighborhood of the Milky Way is not only modified by gravitational forces but 
also by the ambient medium in the gaseous halo and by the strong radiation 
field consisting of stellar UV-photons, soft X-rays and cosmic-rays.

Meyerdierks (\cite{meyerdierks}) detected a HVC that appears like a cometary 
shaped cloud with a central core and an asymmetric envelope of warm neutral atomic 
hydrogen (the particular HVC is denoted in literature as HVC A2). He
interpreted this head-tail structure as the result of an interaction between 
the HVC and normal galactic gas at lower velocities.
Towards HVC complex C Pietz et al. (\cite{pietz96}) discovered the so-called 
\hi ``velocity bridges'' which seem to connect the HVCs with the normal 
rotating interstellar medium.
The most straight forward interpretation for the existence of such
structures is to assume that a fraction of the HVC gas was 
stripped-off from the main condensation.

In the present paper, we extend the investigations of Meyerdierks (\cite{meyerdierks}) 
and Pietz et al. (\cite{pietz96})
over the entire sky covered by the new Leiden/Dwingeloo \hi 21-cm line
survey (henceforward abbreviated as LDS, Hartmann \& Burton \cite{hartmann97}).
For this aim, we investigated the shape and the column density distribution of 
a complete sample of HVCs to search for distortions in the HVC velocity fields
accompanied by column density gradients.

In Sect. 2 we give a brief summary of observational parameters concerning the 
LDS.
In Sect. 3 we present our HVC-sample selection and the characteristic
parameters.
In Sect. 4 we describe the detection process of head-tail structures in our
sample and show the distribution of head-tail structures within the individual 
HVC complexes.
In Sect. 5 we discuss possible implications for the existence of the head-tail
structures.
In Sect. 6 we summarize our results.


\section{The Leiden/Dwingeloo survey}

The Leiden/Dwingeloo survey of Galactic neutral atomic hydrogen (Hartmann \&
Burton \cite{hartmann97}) comprises observations of the entire sky north of 
$\delta$=$-30\degr$. It represents an improvement over earlier large-scale \hi
surveys by an order of magnitude or more in at least one of the principal
parameters of sensitivity, spatial coverage or spectral resolution.
Most important for our scientific aim is the correction of the
LDS for the influence on stray radiation to the \hi spectra (Kalberla et al.
\cite{kalb}, Hartmann et al.
\cite{hartmann96}).
The survey parameters were compiled by Hartmann (\cite{hartmann}) and
Hartmann \& Burton (\cite{hartmann97}). Here we summarize only the major properties 
important for this work.

The angular resolution of the survey is determined by the beam size of the
25-m Dwingeloo telescope to 36'.
The observations were performed on a regular grid with a
true-angle lattice spacing of $0\fdg5$ in both, $l$ and $b$. The velocity 
resolution was set to 1.03 \kmsws per channel of the auto-correlator.
The effective velocity coverage (measured with respect to the
Local Standard of Rest, LSR) covers the range  -450 \kmsws $\leq v_{\rm LSR}
\leq$ 400 \kmsws .
The characteristic RMS limit of the evaluated brightness-temperature
intensities is about 0.07 K.
The residual uncertainties, introduced for instance by baseline fitting, are 
about 0.04 K.


\section{The HVC-sample}
\subsection{Selection criteria}
   
\begin{figure*}[]
  \centerline{
  \hfill
\psfig{figure=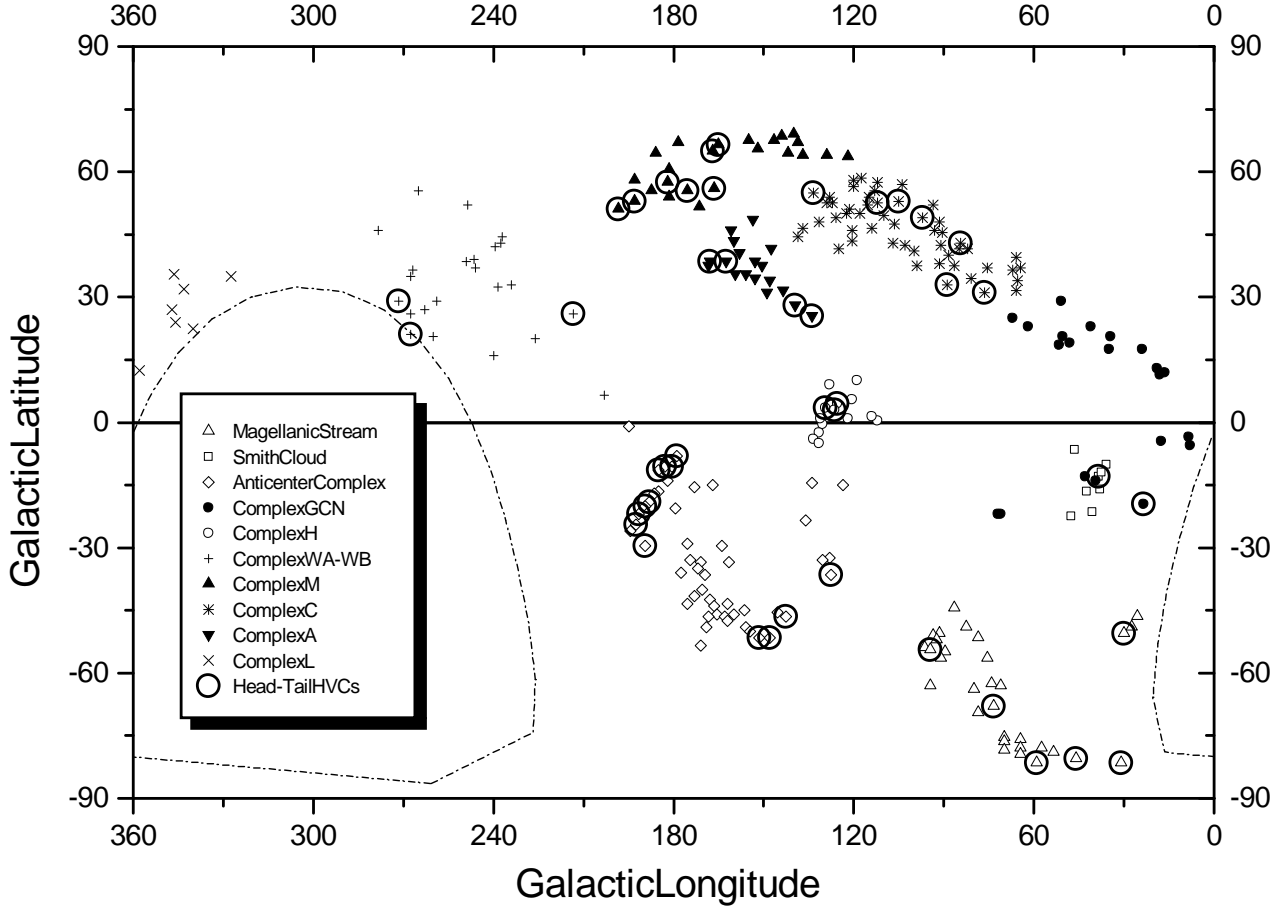,height=12.5cm,angle=0,bbllx=65pt,bblly=460pt,bburx=525pt,bbury=780pt}
  }
 \caption[]{The distribution of the individual HVCs of our sample.
 Each marker represent an individual HVC of our 
  sample. In total 252 HVCs are identified.
  The HVCs showing a head-tail structure are indicated by a superposed open 
  circle. The dashed line indicates the southern limit of the
  LDS at $\delta$ = -30\degr.}
 \label{hvcdist}
\end{figure*}
  
The aim of the present work is to perform a systematic search for
HVCs with a head-tail structure in the entire data base of the
LDS (Br\"uns \cite{bruens}). For a statistical analysis we need a well defined 
and representative sample. Therefore we define three conditions which all 
have to be fulfilled by the HVCs under consideration.  
\begin{itemize} 
\item We identify \hi emission lines as high-velocity \hi profiles if their 
radial velocities are at least $|v_{\rm LSR}| \geq$ 90 \kmsws and deviate at 
least 50 \kmsws~ from a simple galactic rotation model (the second condition is 
important for areas near the Galactic Plane). 
\item The HVC must be traceable within three individual \hi spectra, to
overcome residual baseline uncertainties and broad but faint residual
interference signals. We demand further that the signal is not correlated with
the observational grid. The minimum three \hi spectra are discarded if they 
are aligned only in galactic longitude or latitude. Accordingly, the three 
spectra build up the smallest allowed map of a HVC of interest. This 
corresponds to cloud-sizes larger 1\degr.
\item The minimum \hi column density of the studied HVCs is 
$N_{\rm \hi} = 1\cdot10^{19}$\psqcm. This constraint is introduced to analyze 
only high-velocity \hi profiles with a sufficient signal-to-noise ratio, 
allowing to study the shape of the \hi emission lines in detail. 
\end{itemize}

\noindent Applying the three conditions compiled above to the LDS data, 
we build up a new ``bright source catalogue of extended HVCs'' for the northern sky
offering an angular resolution below $1\degr$.

\subsection{General properties of the HVC sample}
Fig. \ref{hvcdist} shows the distribution of the selected HVCs across the 
galactic sky. 
The dashed-dotted line represents the southern declination boundary of the
LDS at $\delta = -30\degr$. 
Each individual marker represents a single HVC selected according to the three criteria
mentioned above (Sect. 3.1).
In total we identified 252 HVCs.
Because of the applied selection criteria, each marker represents a unique line of
sight to a HVC. 
The selected HVCs follow in detail the positional distribution of the well known
HVC complexes (Wakker \& van Woerden \cite{wakker-rev}).
The open circles mark the location of HVCs with a head-tail structure (see Sect.
4).

 \begin{figure*}[]
  \hfill
  \psfig{figure=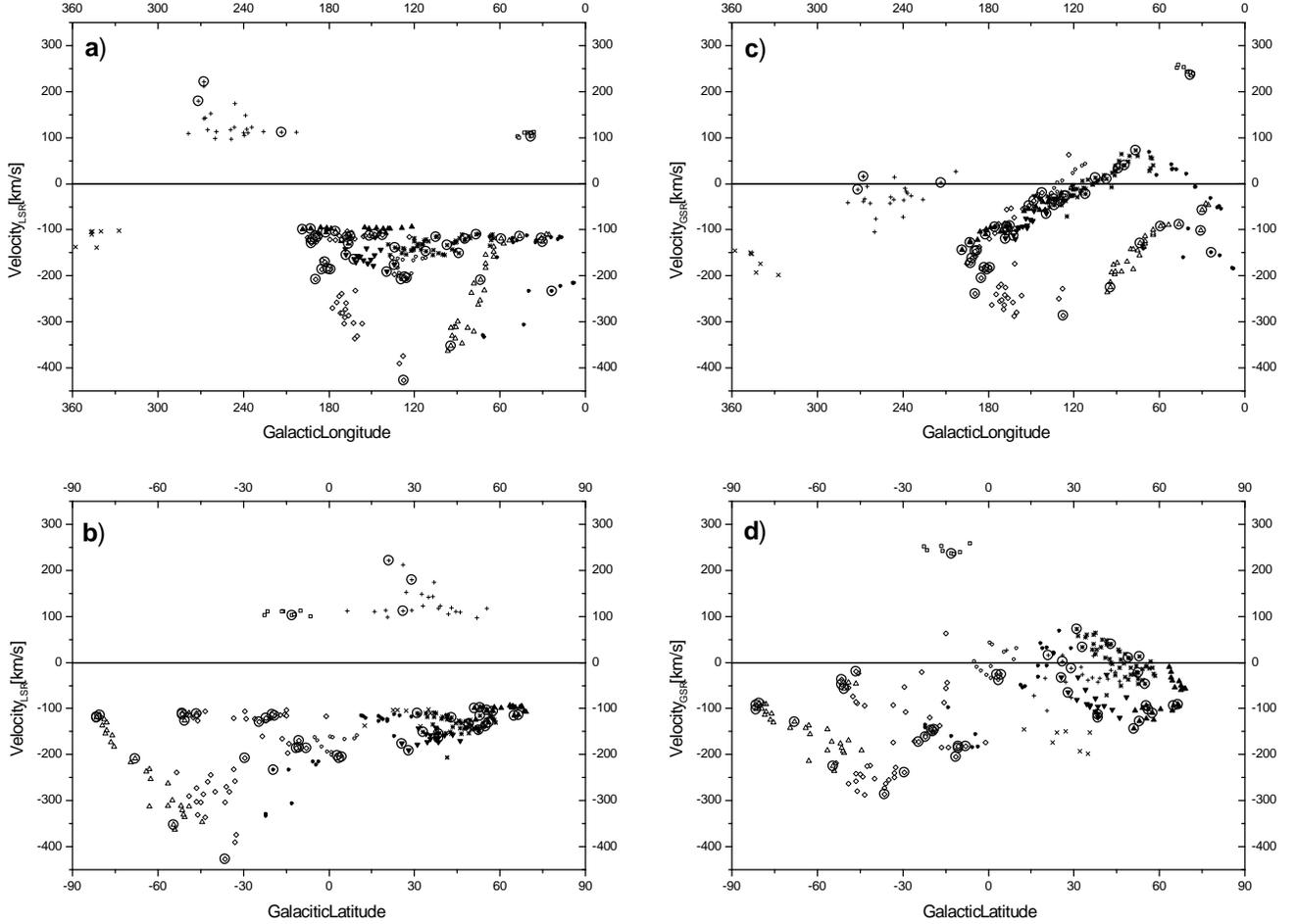,height=13.0cm,angle=0,bbllx=40pt,bblly=410pt,bburx=550pt,bbury=780pt}
  \caption[]{The distribution of the HVCs - radial velocity (in $v_{\rm LSR}$ 
  and $v_{\rm GSR}$) as a function of the parameters galactic
   longitude and latitude:
   \textbf{a}  $v_{\rm LSR}$ vs. galactic longitude, 
   \textbf{b} $v_{\rm LSR}$ vs. galactic latitude, 
   \textbf{c} $v_{\rm GSR}$ vs. galactic longitude, 
   \textbf{d} $v_{\rm GSR}$ vs. galactic latitude. The
   transformation from LSR to GSR was calculated according Eq. 1. 
   The HVCs showing a head-tail structure are marked (as in Fig. 1) 
   with open circles.}
  \label{param}
  \hfill
 \end{figure*}

 \begin{figure*}[]
  \centerline{
  \hfill
  \psfig{figure=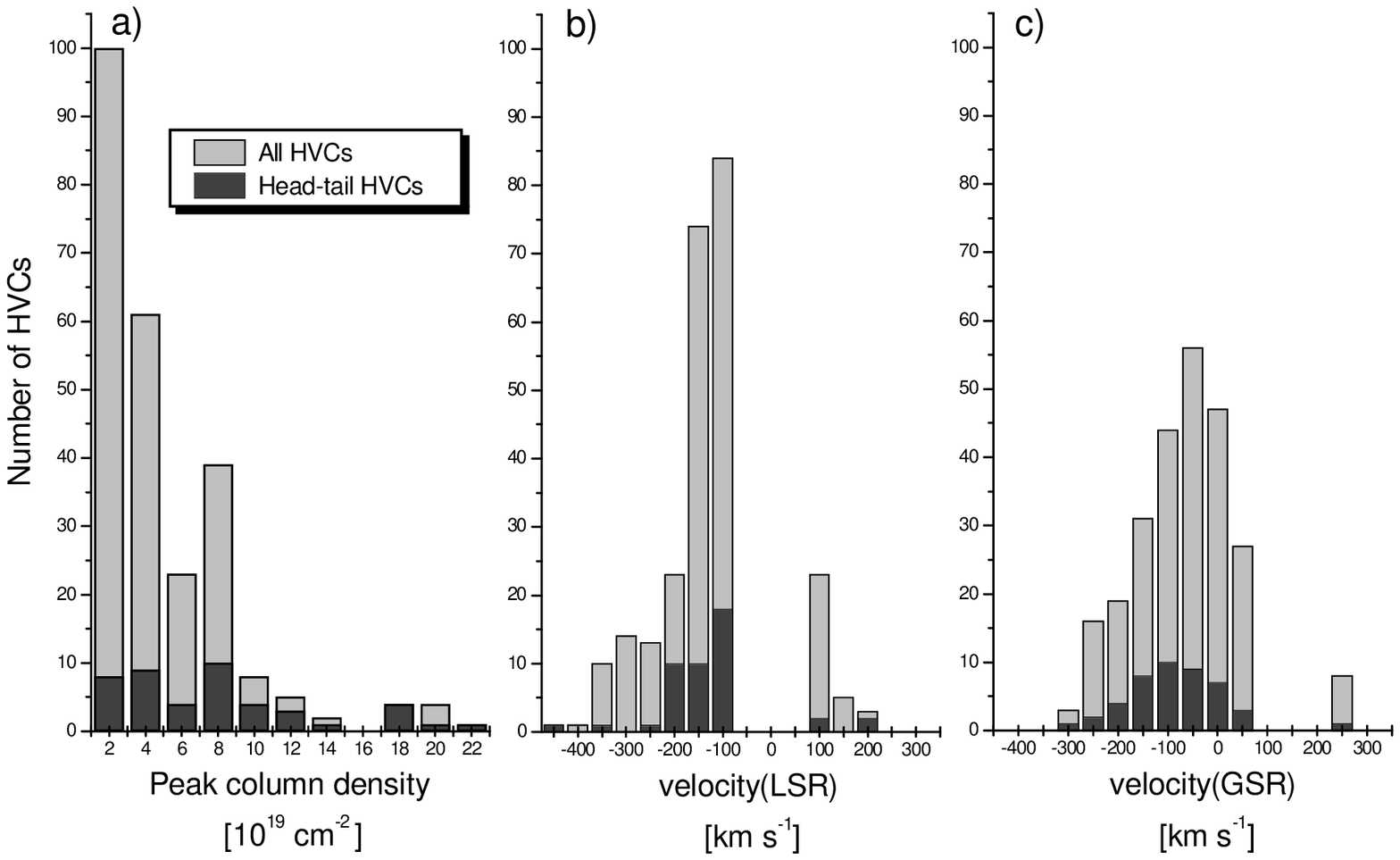,height=10.8cm,angle=0,bbllx=30pt,bblly=470pt,bburx=560pt,bbury=800pt}
  }
  \caption[]{These histograms show the number of HVCs versus \textbf{a} peak column 
  density, \textbf{b} $v_{\rm LSR}$ and \textbf{c} $v_{\rm GSR}$.
  The entire HVC sample is plotted in light gray, the HT HVCs are plotted in 
  dark gray. ($\Sigma_{HVC}$ = 252, $\Sigma_{head-tail}$ = 45)}
  \label{histograms}
 \end{figure*}
 
In Fig. \ref{param}a the radial velocity (in the local standard-of-rest frame) 
is plotted versus the galactic longitude. 
Most of the HVCs have velocities in the range of -200 \kmsws $\leq v_{\rm LSR} \leq$
-100 \kmsws or +100 \kmsws $\le v_{\rm LSR} \leq$ +200 \kmsws.
Only HVCs belonging to the Magellanic Stream, the anti-center
 complex and some clouds near the Galactic Center show radial 
velocities of v$_{\rm LSR} \leq$ -200 \kmsws.
All these more extreme velocity HVCs are located on the southern galactic sky.
The absence of a significant number of HVCs with positive radial velocities is because
of the limitation to $\delta \geq -30\degr$ of the LDS data.
In particular the Magellanic Cloud System is traceable in \hi 21-cm line radiation
with positive $v_{\rm LSR}$ velocities (see Putman et al. \cite{putman} or
Putman \cite{putman99} for a recent overview).

Fig. \ref{param}b shows the radial velocity (in the local standard of rest)
as a function of the galactic latitude. 
In this figure the north/south asymmetry is visible, i.e. the HVCs in the 
northern galactic sky have relatively low radial velocities while the HVCs on 
the southern galactic sky show in general much higher radial velocities.

Figs. \ref{param}c and d show our HVC sample in a different representation:
radial velocities are transformed into the galactic standard-of-rest frame.
\begin{eqnarray}
v_{\rm GSR} = v_{\rm LSR} + 220~{\rm sin}(l){\rm cos}(b)
\label{vgsr}
\end{eqnarray}
In Fig. \ref{param}c the radial velocity (v$_{\rm GSR}$) is plotted versus the
galactic longitude. The HVC complexes are grouped now to coherent structures, 
which are even larger than the extent of an individual HVC complex.
For example, the HVC complexes M, A and C build up the largest coherent structure in 
this representation.
Most of the HT HVCs belong to this structure.
The Magellanic Stream forms a parallel shifted feature.
There is one cloud complex that is clearly outside the main distribution: the 
Smith cloud (l = 38\degr, b = --13\degr, Smith \cite{smith}) shows radial velocities 
v$_{\rm GSR} \approx$ +250 \kmsws while all other clouds have 
v$_{\rm GSR} \leq$ +75 \kmsws. This may be a hint for a different origin of the
Smith cloud. 
Bland-Hawthorn et al. (\cite{bland-hawthorn}) claimed an association of the
Smith cloud with the Sgr dwarf.\\
Fig. \ref{param}d shows the radial
velocity (v$_{\rm GSR}$) as a function of the galactic latitude. 
The north/south asymmetry is still visible.

Fig. \ref{histograms} shows histograms of the HVC distribution for the 
parameters peak column density, $v_{\rm LSR}$ and $v_{\rm GSR}$.
The histogram of peak column density versus number distribution of the corresponding HVCs
contain information on the studied ensemble.
We evaluated the log($N$)-log($S$) correlation (where $N$ is the number of HVCs
per flux interval and $S$ is the flux) and found a correlation
coefficient of -0.91, clearly indicating a linear relation between both 
quantities.
The linear equation is {log($N$) = $(0.53\,\pm\,0.07)\times$ log($S$)\,+\,
$(3.00\,\pm\,0.06)$.
Within the uncertainties these numbers are equal to the values of Wakker \& van 
Woerden (\cite{wakker91}) on the log($N$)-log($S$) of the population of 
positive and negative HVCs distributed across the northern sky.
This result demonstrates, that our ensemble of HVCs is a representative
flux-limited sample of HVCs. \\
Fig. \ref{histograms}b shows the number distribution with respect to the 
radial velocity ($v_{\rm LSR}$). Most of the HVCs have velocities in the
intervals centered on $v_{\rm LSR}$ = --150 \kmsws and --100 \kmsws.  
Fig. \ref{histograms}c shows the number distribution with respect to the 
velocity ($v_{\rm GSR}$).
The bulk of the studied HVCs reveal low radial velocities with respect
to the galactic standard of rest frame.
The mean $v_{\rm GSR}$ velocity is negative ($\overline{v_{\rm HVC}}$ = --62.3
\kmsws $\pm$ 7.5 \kmsws), indicating that the majority of
HVCs in our sample are moving towards the Galactic Disk.
Wakker \& van Woerden (\cite{wakker91}) showed that the inclusion of the southern hemisphere \hi
data on HVCs in their sample does not change this general $v_{\rm GSR}$ velocity
behavior. \\


\section{Head-Tail structures}

Our HVC sample, selected according to the conditions compiled in Sect. 3,
provide information on the distribution of the HVCs relative to the 
observational parameters galactic longitude and latitude, radial velocity
($v_{\rm LSR},~v_{\rm GSR}$) and peak column density.

In the following step we analyze each individual HVC of the sample with
respect to the shape of its \hi line profiles, for instance for the variation 
of the mean velocity and the variation of the column density across the HVC
extent. 

     \begin{figure*}
       \centerline{
       \hfill
       \includegraphics{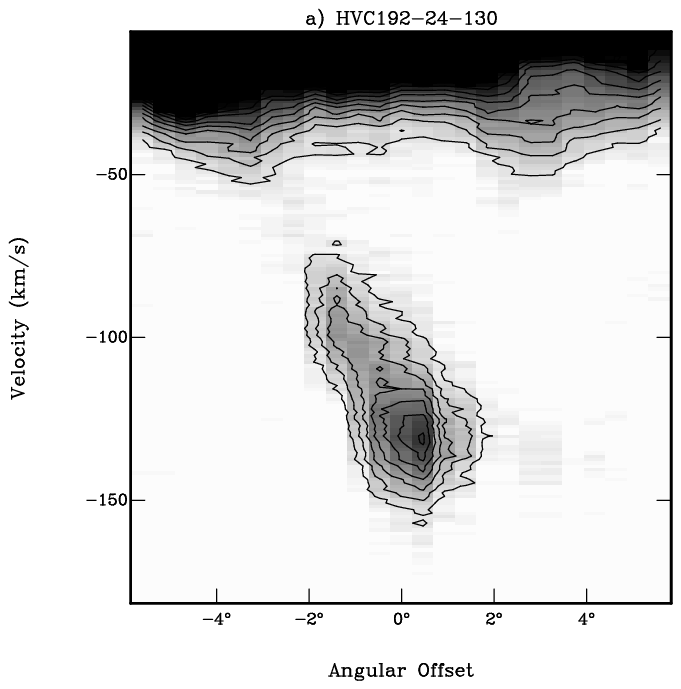}
       \hfill
       \includegraphics{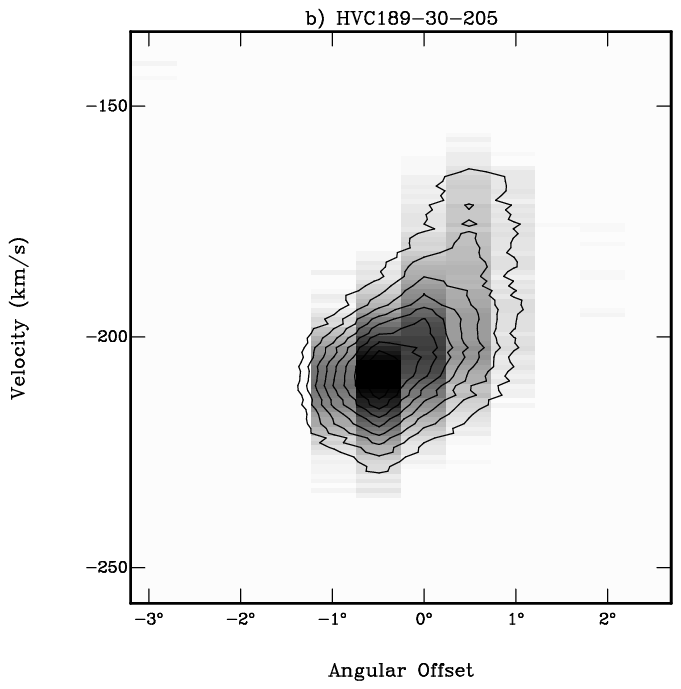}
       \hfill
       } 
       \centerline{
       \hfill
       \includegraphics{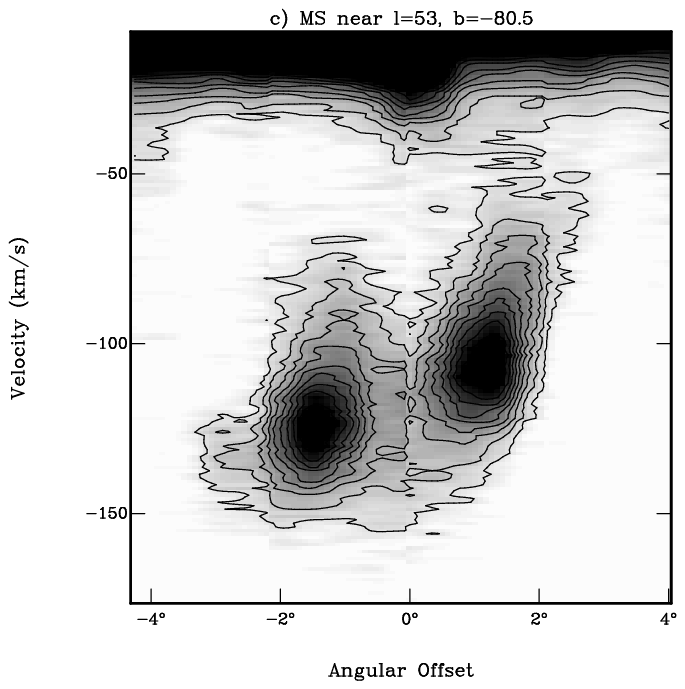}
       \hfill
       \includegraphics{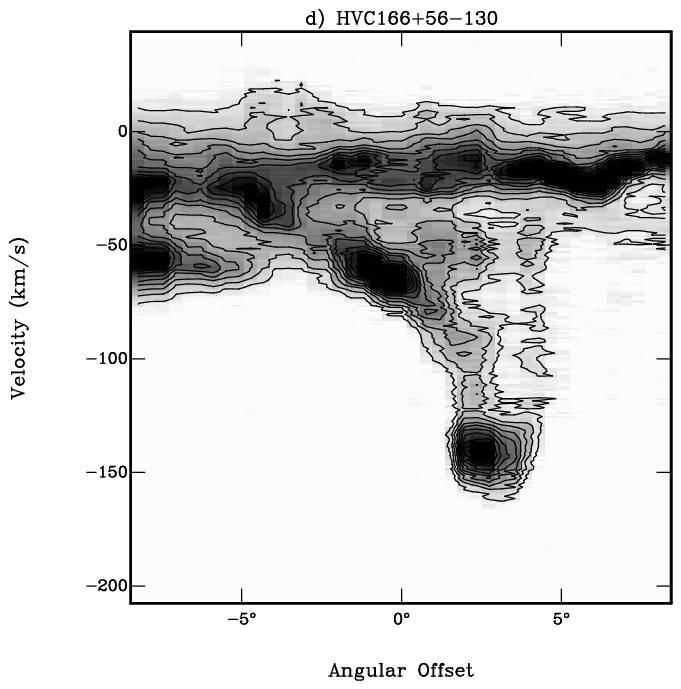}
       \hfill
       }
       \centerline{
       \hfill
       \includegraphics{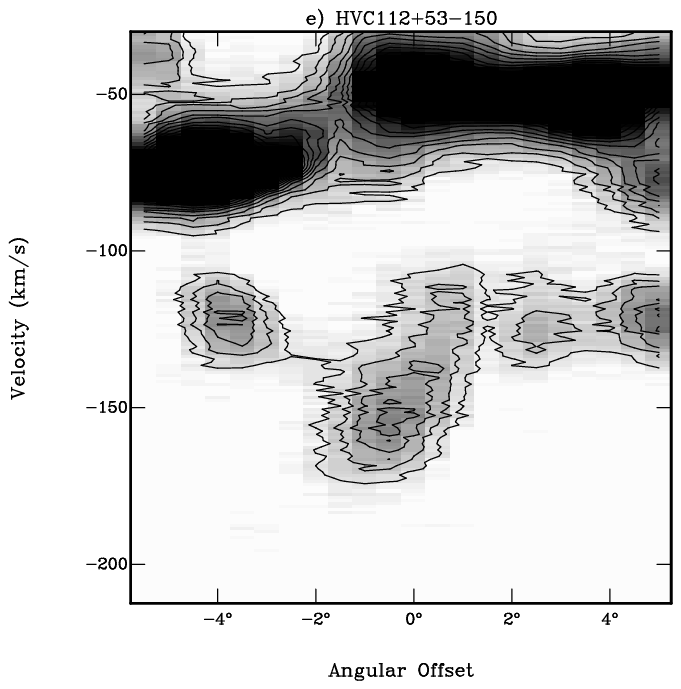}
       \hfill
       \includegraphics{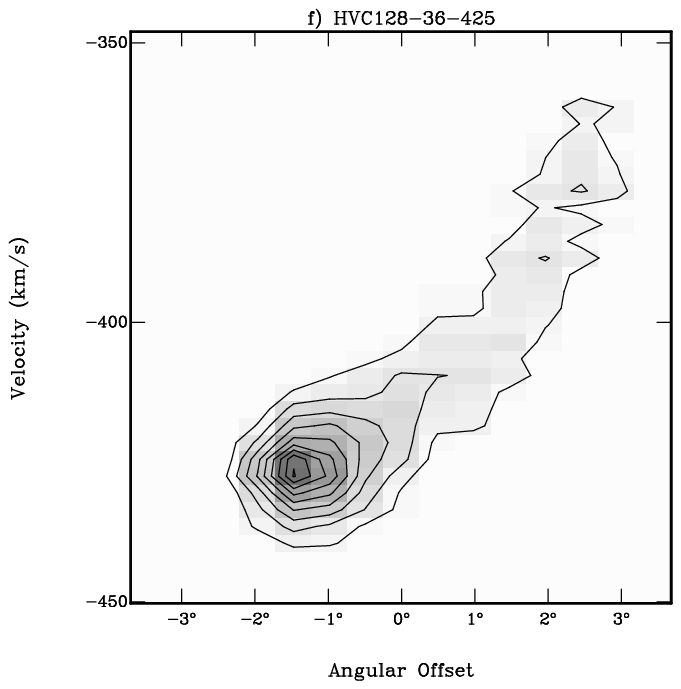}
       \hfill
       }
       \caption[]{Examples of HVCs with a head-tail structure. These
       position-velocity diagrams are oriented along the velocity 
       gradient axis of each head-tail HVC: \textbf{a} HVC192-24-130, 
       \textbf{b} HVC189-30-205,
       \textbf{c} two clouds in the Magellanic Stream (near l=53\degr, b=--80\degr.5,
       \textbf{d} HVC166+56-130 \textbf{e} HVC112+53-150 and \textbf{f} HVC128-36-425; 
       the contour lines start at T$_{\rm B}$ = 0.21 K (3$\sigma$) in steps of 
       $\Delta$T$_{\rm B}$ = 0.14 K(2$\sigma$), gray-scale indicates T$_B$, 
       the darker the color the higher T$_{\rm B}$ (each Fig. has the same
       intensity scale). 
       The contour lines in \textbf{f} start at T$_{\rm B}$ = 0.12 K (3$\sigma$ at 
       $\Delta$v = 3\kmsws) in steps of $\Delta$T$_{\rm B}$ = 0.08 K(2$\sigma$).
       } 
       \label{headtailpic}
   \end{figure*}
   
\subsection{Velocity gradients}
   
We find that about 40\% of the HVCs show up with a significant velocity 
gradient.
These velocity gradients are very frequently associated with a column 
density gradient (20\%).
The detected velocity gradients are by them-self not {\em a priori} 
indicators for an intrinsic distortion of the HVC velocity field.
There are several effects that can produce a velocity gradient.

\begin{itemize}
   
\item The HVCs of our sample are, because of the applied selection criteria, 
extended objects.
Accordingly, the angle between the line of sight and the solar velocity vector 
varies across the extent of the HVC. This effect can produce a velocity 
gradient of about 10~{\kmsws}/${\rm [ \degr ]}$.
\item The same kind of velocity gradient is also expected from the HVC velocity
vector, because the angle between the HVC velocity vector and the line of sight 
changes with position, too. This velocity gradient depends on the unknown 3D 
velocity of a HVC.   
\item Two or more HVCs may be superposed on the same line of sight with 
comparable group velocities. The probability for an accidental superposition 
of two independent HVCs is very low. Nevertheless it is known that
HVCs consist of several clumps. The superposition of two clumps that belong to
the same HVC has a much higher probability. High angular resolution \hi 
studies may disclose this kind of arrangement.
\item Even if there is no evidence for a rotating HVC so far, a rotating HVC
would show up with a velocity gradient similar to rotation curves of galaxies.
Revealing a red- and blue-shifted extension in the position-velocity diagram.
\end{itemize} 
The effect related to the solar velocity vector is well defined and therefore
easy to calculate according to Eq. \ref{vgsr}.
All other velocity gradients are related to the HVC phenomenon.

\subsection{Definition of head-tail structures}

Our aim was to search for real distortions in the velocity field of
individual HVCs, which may be an indicator for an interaction of the HVC
with the surrounding interstellar medium.
Accordingly, we studied those HVCs which reveal a velocity {\em and} a column
density gradient simultaneously. 
This kind of HVC appears like a comet in the position-velocity domain, and
justifies the name head-tail HVC (HT HVC).
The head is the region with the highest column density of the HVC, while the
column density of the tail is in general much lower (by a factor of 2 - 4). 
Fig. \ref{headtailpic} shows six examples of the studied HT HVCs:
the HVC shown in Fig. \ref{headtailpic}a and b 
belong to the anti-center complex (HVC189-30-205 and HVC192-24-130). 
In Fig. \ref{headtailpic}c two HVCs of the
Magellanic Stream are displayed. Both show a head-tail structure. 
The HVC displayed in Fig. \ref{headtailpic}d (HVC166+56-130) is a special 
HT HVC, because the HVC appears to be connected with the \hi gas in the
Galactic Disk; it forms a so called velocity bridge.
Up to now it is unknown, whether the HVC gas is really physically connected to
the Galactic Disk gas or just an accidental superposition on the same
line of sight.
Fig. \ref{headtailpic}e shows a HT HVC where the head has a relatively low
intensity.
Fig. \ref{headtailpic}f shows the HT HVC with the highest observed radial
velocity in our sample (v$_{\rm LSR}$ = --425 \kmsws). The velocity resolution 
was reduced to 3 \kmsws in this picture to give an idea of the 
extent of the tail. This HVC is a very-high-velocity cloud (VHVC) several 
degrees away from the galaxy M33 (--300 \kmsws $\le$ v$_{\rm LSR}(\rm M33) \le$
--75 \kmsws). 

    \begin{figure}
       \centerline{
       \hfill
       \psfig{figure=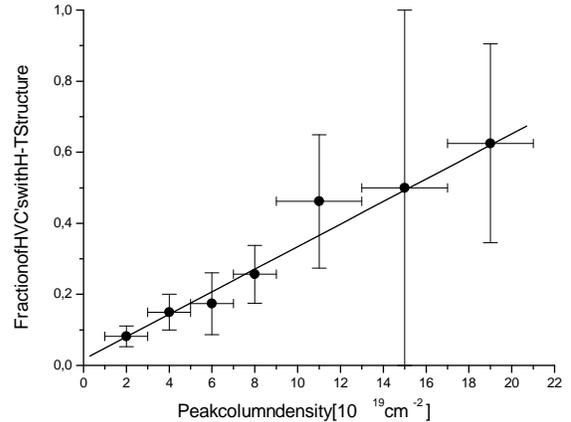,height=6cm,angle=0,bbllx=75pt,bblly=445pt,bburx=510pt,bbury=770pt}
       \hfill
       } 
       \caption[]{Correlation between the fraction of HT HVCs on all HVCs of the sample
       in a certain column density interval versus the peak column density. The
       solid line marks the result of a weighted linear regression.} 
       \label{hvc-linear} 
   \end{figure}

\subsection{Results}

In total, we found 45 HVCs associated with a HT structure in our HVC-sample 
containing 252 individual HVCs.
The column density contrast between the head and the tail varies in general 
between factors 2 -- 4. It is a general behavior that the column density 
maximum, the head, is located at higher radial velocities than the tail. The HT HVCs 
are extended objects and cover in general several square degrees.

Fig. \ref{hvcdist} shows the distribution of the head-tail HVCs across
the galactic sky. The HT HVCs are marked by open circles. Obviously all HVC 
complexes reveal HT HVCs, even in the northern part of the Magellanic Stream 
HT HVCs are present. Only in HVC complex L no HT HVCs were found. 
Fig. \ref{param} shows the distribution of the HT HVCs in comparison to the
HVC sample in respect to the parameters galactic longitude, latitude and radial
velocity ($v_{\rm LSR}$ and $v_{\rm GSR}$).

In Fig. \ref{histograms} histograms are plotted for the whole HVC sample and
the HT HVCs with respect to the parameters peak column density, $v_{\rm LSR}$ and 
$v_{\rm GSR}$. We discuss these histograms further in the following subsections.

\subsubsection{Peak column density}\label{lin}

Fig. \ref{histograms}a shows histograms for the number of HVCs per column
density interval. The histogram for the entire HVC sample is plotted in light 
gray, the histogram for the HT HVCs is plotted in dark gray.
It is obvious that there are only relatively few HT HVCs at lower column 
densities (relative to the HVC sample).\\
A detailed analysis of the histogram for the peak column density revealed
a linear relation between the fraction of head-tail HVCs and the peak 
column density. Fig. \ref{hvc-linear} shows this relation. The vertical 
error-bars are calculated with respect to the low statistics (there are very 
few HT HVCs with a high column density), the horizontal error-bars indicate 
the size of the individual $N_{\rm \hi}$ intervals. 
The solid line represents the linear regression fit:
\begin{eqnarray}
\frac{\#HTs}{\#HVCs} = (0.0317 \pm 0.0028) \frac{N(\hi)}{[10^{19}
{\rm cm^{-2}}]} +\nonumber\\ 
+ (0.017 \pm 0.012)
\label{eqlin}
\end{eqnarray}
We like to emphasize that the lower limit of $N_{\rm \hi} = 1~10^{19}$\psqcm 
was chosen to derive statistical significant information on even 
the faintest HVCs in our sample (this limit is about a factor of 3 above the 
detection limit using the highest velocity resolution $\Delta$v = 1.03 \kmsws of
the LDS data). The correlation displayed in Fig. \ref{hvc-linear} is {\em not} 
biased by a detection limit.\\
This empirical relation can be used to 
determine an expected number of HT HVCs
for each individual HVC-complex in our sample. The comparison of the expected 
numbers of HT HVCs with the detected ones shows a good consistency.
Only towards the Galactic Anti-Center a significantly larger number of 
HT HVCs is detected than expected. 
The correlation explains why we were not able to find HT HVCs in Complex L: this
complex has only very few, faint HVCs (the expected number of HT HVCs for this
complex is only 0.3).

\subsubsection{Radial velocity ($v_{\rm LSR}$)}

Fig. \ref{histograms}b shows histograms for the number of HVCs per velocity
($v_{\rm LSR}$) interval. The histogram for the entire HVC sample is plotted in light 
gray, the histogram for the HT HVCs is plotted in dark gray.\\
A large number of HT HVCs have radial velocities in the intervals centered on 
--100 and --150 \kmsws (62.2\%). This is consistent with the fact that most HVCs
have radial velocities in this regime (62.9\%).
The interval centered on --200 \kmsws contains a relatively large fraction of HT
HVCs. The large number is expected because of the fact that there is a high
percentage of high column density HVCs in this velocity interval, increasing the
probability to detect a HT HVC according to Eq. 2.
It remains unclear whether there are too many high column density HVCs or too
few low column density HVCs in this velocity interval.

\subsubsection{Radial velocity ($v_{\rm GSR}$)}

Fig. \ref{histograms}c shows histograms for the number of HVCs per velocity
($v_{\rm GSR}$) interval. The histogram for the entire HVC sample is plotted in light 
gray, the histogram for the HT HVCs is plotted in dark gray.\\
The mean radial velocity $v_{\rm GSR}$ of the HT HVCs is more
negative than the mean radial velocity of the HVC sample. A Gaussian fit to the
histograms revealed mean velocities of $\overline{v_{\rm HT HVC}}$ = --86.1 
\kmsws $\pm$ 3.8 \kmsws and 
$\overline{v_{\rm all HVC}}$ = --62.3 \kmsws $\pm$ 7.5 \kmsws.
The difference between the mean velocities ($\Delta$v=23.8\kmsws) is
significant. 
The probability to find a HT HVC increases with negative 
GSR velocity.
We checked that this result is {\em not} caused by the relation shown in 
Fig. \ref{hvc-linear}.\\
The velocity dispersion in Fig. \ref{histograms}c is nearly identical for the 
HT HVCs (FWHM = 174.0 \kmsws $\pm$ 7.5 \kmsws) and the complete HVC sample 
(FWHM = 179.9 \kmsws $\pm$ 15.1 \kmsws).


\section{Discussion}
In the following subsections we discuss possible interpretations for
HVCs with a head-tail structure at different distances from the Galactic Disk.

\subsection{HVCs in the gaseous Galactic halo}

The gaseous halo of the Milky Way has a vertical scale height of 4.4 kpc
(Kalberla \& Kerp \cite{kalberla}). At least the HVC complexes M (z $\leq$
3.5 kpc, Danly et al. \cite{danly}, Keenan et al. (\cite{keenan}) and Ryans
et al. (\cite{ryans})) and A (2.5 $\leq$ z $\leq$ 7 kpc,
van Woerden et al. \cite{van Woerden}) are located within the gaseous halo.
As a natural consequence some kind of interaction is expected when a
cloud has a high velocity relative to the ambient medium. In a ram-pressure
model the head is the interacting HVC and the tail is gas which was recently
stripped off from the HVC.
In an absolutely homogeneous halo medium there should be a constant interaction
rate for
all HVCs with similar distances and velocities. The HVC complexes form coherent
structures in position and in velocity, i.e. they may form also coherent
structures in space. Therefore one would expect naively that all HVCs of a 
complex should show up with a
head-tail structure. This is obviously not observed. \\
Kalberla \& Kerp \cite{kalberla} estimate that $\approx$ 10 \% of the halo gas
is neutral, having higher density than the surrounding plasma. 
This implies that only those HVCs are expected to show up with a head-tail
structure that are passing through an area with locally higher halo density. 
Accordingly there is a 10 \% probability for the creation of a HT HVC. 
Thus at least 10 \% of the HVCs should have a head-tail structure.

In the following we estimate the life-time of a head-tail structure.
The HVC tails have on the average a column density of a few times 
$10^{19}$ \psqcm. Because of the low dust abundance in galactic HVCs 
(in particular in case of the galactic HVC complex M, Wakker \& Boulanger 
\cite{wakker86}) photoelectric heating can be neglected as an excitation process.
Only the diffuse X-ray radiation of the galactic halo and the cosmic-rays heat
the HVCs.
According to Wolfire et al. (\cite{wolfirea}a and b) column densities of a few $N_{\rm \hi} \sim 
10^{19}$ \psqcm have a high ionization probability.
This corresponds to a life-time in the neutral state of about 
$10^5$ -- $10^6$ years. Compared to the free-fall time 
of a galactic HVC, of about $10^7$ years, the life-time of the stripped-off
matter is very short. The tail of a high peak column density HVC contains more
material and will survive significantly longer against the ionizing radiation 
of the diffuse X-rays and cosmic rays.
This is consistent with the observational fact that the high peak column 
density HVCs reveal much more frequently a HT HVC (Fig. \ref{hvc-linear}).\\
In addition, it is possible that the production or the life-time of a tail
depends on the small scale structure of a HVC. HVCs consist of two components:
cold clumps are surrounded by an envelope of warm gas. HT HVCs with clumps in
their tails will survive much longer than tails that contain only diffuse warm
gas. The angular resolution of the LDS is too low to resolve the inner parts 
of the HVCs. \hi observations with high angular resolution are mandatory to 
study the effects of the small-scale structure.\\

\subsection{The Magellanic Stream}

The Magellanic Stream (MS) is the only HVC complex with a generally accepted origin:
it is build up by gaseous debris caused by the tidal interaction between the Magellanic Clouds and the
Milky Way. The distance of the southern end of the Magellanic Stream is assumed
to be very similar to the distance to the Magellanic Clouds (D $\approx$ 50 kpc). In
the opposite, the distance of the northern end is unknown.
If it is located at low z-heights, i.e. in the gaseous halo, the observed
head-tail structures may be produced by the same process as the HT HVCs in the
complexes M and A. In this scenario the MS covers a large distance bracket  of
several tens of kpc. There should be no head-tail structures
in the southern part of the stream.
On the other hand, the northern part of the MS could have distances similar to
the distance of the Magellanic Clouds. The existence of HT HVCs
cannot be explained by an interaction with the gaseous halo, because it would be
far outside the gaseous halo. A possible interaction partner could be the debris
from previous revolutions of the Magellanic System. The ionizing radiation field 
is much weaker at these distances. Accordingly, the life-time of a tail in the 
MS is orders of magnitudes longer compared to HVCs in the lower Galactic halo.  

\subsection{HVCs at extragalactic distances}

There is evidence that some HVCs are at intergalactic distances
(Braun \& Burton \cite{braun}, Blitz et al. \cite{blitz}, 
Sembach et al. \cite{sembach}).\\
Sembach et al. (\cite{sembach}) found highly ionized high-velocity gas clouds.
Several lines of evidence, including very low thermal pressures (P/k $\approx$ 2
cm$^{-3}$ K), favor a location for the highly ionised high-velocity gas clouds
in the Local Group or very distant Galactic halo.\\
Braun \& Burton 
(\cite{braun}) searched for compact, isolated HVCs. They classified HVCs as 
compact if they have angular sizes less than 2 degrees FWHM. They are isolated 
in that sense, that they are are separated from neighboring \hi emission by 
expanses where no emission is seen to the detection limit of the data. 
They found 66 of these compact, isolated HVCs.
A comparison between their and our sample reveal that the sample of Braun \& 
Burton is much more uniformly distributed in the parameter space (position vs. 
velocity) than the HVCs of our sample. The different distributions may be an
indication for different objects in origin and evolution. 
Braun \& Burton claimed, that their HVCs are most likely located at 
intergalactic distances. The HVCs of our sample probably contain a  
significant number of HVCs located
in the gaseous halo of the Milky Way. 11 HVCs of their sample are also included in 
our sample. Moreover, two of them show up with a head-tail structure 
(HVC271+29+181 and HVC30-51-119).\\
If these HVCs are located in the intergalactic space, the life-time of the
head-tail structures would be much longer because of the much weaker ionizing
radiation field at these distances.
On the other hand, it is not straight forward to explain what kind of interaction process
may produce the observed features at distances of several hundreds of kpc.
Especially these (probably very distant) head-tail HVCs should be observed with a
higher angular resolution in the near future. 


\section{Summary and conclusion}
We performed a systematic analysis of the HT HVCs across the whole sky 
that is covered by the LDS (about 75\% of the entire sky).
We selected a representative flux limited HVC sample with column densities
$N_{\rm \hi} \geq 1~10^{19}$\psqcm and minimum angular diameter $\geq$ 1\degr.
In total we found 252 HVCs.

Each individual HVC was analyzed with respect to velocity gradients 
and asymmetries in the \hi line profiles. The so called head tail HVCs 
have a cometary shape in the position-velocity domain.
We found that 45 out of 252 HVCs show up with a head-tail structure.
These HT HVCs are randomly distributed over the whole sky covered by the LDS.

A statistical evaluation of the HVC ensemble revealed that the probability to 
detect a HT structure increases linear with the peak column density and with
increasing negative radial velocity in the GSR frame.
There is no significant correlation between the other parameters like
galactic longitude, latitude and $v_\mathrm{LSR}$.

The detection of HT HVCs in nearly all prominent HVC complexes implies
qualitatively comparable physical processes in all of the HVC complexes.
Individual HVC cores seem to interact with their ambient medium.
In case of the HVC complexes located within the Galactic halo and the 
Magellanic Stream the interaction with the gaseous halo medium or gaseous 
debris distributed along the orbit of the Magellanic Clouds may serve as a 
straight forward explanation for the existence  of the HT HVC.
At intergalactic distances the detection of HT HVCs is difficult to interpret,
because the characteristic time-scale for heating and cooling of the HVC matter
is orders of magnitude shorter than the assumed age of these HVC complexes.

High angular resolution \hi observations of the detected 45 HT HVCs in future 
will improve our knowledge on the temperature structure and small-scale column 
density distribution of this special kind of HVCs. 
In addition the comparison with other wavelengths, e.g. with \halpha emission,
will help to understand the existence of HT HVCs.


\begin{acknowledgements}
      Part of this work was supported by the German
      \emph{Deut\-sche For\-schungs\-ge\-mein\-schaft, DFG\/} project
      number ME 745/19.
\end{acknowledgements}

\end{document}